\documentclass[runningheads]{llncs}

\usepackage{graphicx}
\usepackage[numbers]{natbib}
\usepackage{amsmath}
\usepackage{amssymb}
\usepackage{caption}
\usepackage{tabto}
\usepackage{csquotes}
\usepackage{enumitem}
\usepackage{mathtools}
\usepackage{bm}
\usepackage{environ}
\usepackage{etoolbox}
\usepackage{fancyvrb}
\usepackage{ifthen}
\usepackage{kvoptions}
\usepackage{catchfile}
\usepackage{listings}
\usepackage[bibliography=common]{apxproof}

\newcommand{\term}[1]{\textbf{#1}}

\newcommand{\True}{\textsc{True}}

\newcommand{\Remainder}{\texttt{Remainder}}
\newcommand{\Threshold}{\texttt{Threshold}}

\newcommand{\Or}{\texttt{Or}}

\newcommand{\protocol}[1]{\textsc{#1}}

\newcommand{\Initiator}{\textsf{Initiator}}
\newcommand{\Responder}{\textsf{Responder}}

\newcommand{\Sender}{\textsf{Sender}}
\newcommand{\Receiver}{\textsf{Receiver}}
\newcommand{\view}{\textsf{view}}
\newcommand{\bview}{\textbf{\textit{\textsf{view}}}}

\newcommand{\A}{\mathcal{A}}
\newcommand{\mf}[1]{\mathfrak{#1}}
\newcommand{\omf}[1]{\overline{\mathfrak{#1}}}
\newcommand{\multiset}{\text{\upshape{multiset}}}

\newtheoremrep{theorem}{Theorem}

\newcommand{\emptysec}{\subsubsection*{}

\vspace{-14pt}

\hspace{-6pt}}

\begin{document}

\title{Privacy in Population Protocols with Probabilistic Scheduling}

\author{Talley Amir \and James Aspnes}
\authorrunning{T. Amir \and J. Aspnes}
\institute{Yale University, New Haven CT 06511, USA\\
\email{\{talley.amir,james.aspnes\}@yale.edu}}

\maketitle

\begin{abstract}
The population protocol model \cite{AADFP2006} offers a theoretical framework for designing and analyzing distributed algorithms among limited-resource mobile agents. While the original population protocol model considers the concept of anonymity, the issue of privacy is not investigated thoroughly. However, there is a need for time- and space-efficient privacy-preserving techniques in the population protocol model if these algorithms are to be implemented in settings handling sensitive data, such as sensor networks, IoT devices, and drones. In this work, we introduce several formal definitions of privacy, ranging from assuring only plausible deniability of the population input vector to having a full information-theoretic guarantee that knowledge beyond an agent's input and output bear no influence on the probability of a particular input vector. We then apply these definitions to both existing and novel protocols. We show that the \Remainder-computing protocol from \cite{DFGR2007} (which is proven to satisfy output independent privacy under adversarial scheduling) is not information-theoretically private under probabilistic scheduling. In contrast, we provide a new algorithm and demonstrate that it correctly and information-theoretically privately computes \Remainder\ under probabilistic scheduling.
\end{abstract}
\section{Introduction}

Various issues arise when applying the theoretical population protocol model to real-world systems, one of the most critical of which is that of preserving privacy. The motivation for furthering the study of privacy within population protocols is to better adapt these algorithms to the real-world systems that they aim to model, such as sensor networks, systems of IoT devices, and swarms of drones, all of which handle sensitive data. Previous research in private population protocols only considers adversarial scheduling, which makes generous assumptions about our obliviousness to the scheduler's interaction choices and offers only very weak criteria for satisfying the definition of ``privacy.'' In this work, we further refine these definitions considering a realistic range of threat models and security concerns under arbitrary schedules.
\subsection{Related Work}

Research in private computation within ad hoc networks is distributed (pun intended) over multiple academic fields. We limit our review of the literature to works that most closely relate to the theoretical model we study in this paper.

\subsubsection{Population Protocols}

Privacy was introduced to the population protocol model in \cite{DFGR2007}, where the authors 
define a notion of privacy called \textit{output independent privacy} and provide protocols satisfying this definition for computing the semilinear predicates. Output independent privacy basically states that for any input vector and execution yielding a particular sequence of observations at an agent, there exists a different input vector and execution yielding the same sequence of observations at that agent. The practicality of this definition relies on adversarial scheduling, which allows the schedule of interactions to delay pairs of agents from interacting for an unbounded number of steps. Due to adversarial scheduling, the \textit{existence} of an execution is sufficient to achieve plausible deniability: Agents have no metric for estimating time elapsed nor approximating how many interactions in which another agent has participated. Therefore, the observed state of an agent cannot be used to infer the agent's input as it may have deviated from its original state over the course of many interactions. However, if instead the scheduler is probabilistic, then there arises the issue of data leakage from inferring the population's interaction patterns.

\subsubsection{Sensor Networks}

Population protocols are designed to model sensor networks, but there is a large body of literature on sensor networks that is not connected to the population protocol model. 
The capacity of agents in the domain of sensor networks is much larger than is assumed in population protocols; in particular, much of the privacy-preserving algorithms in this area involve encryption, which requires linear state space in the size of the population.

In recent years, viral exposure notification via Bluetooth has become a popular area of study \cite{Canetti2020PrivacyPreservingAE, Chan2020Pact}, and one that demands verifiable privacy guarantees due to widespread laws governing protected health data. However, the solutions in \cite{Canetti2020PrivacyPreservingAE, Chan2020Pact} require centralization and high storage overhead. The closely related problem of anonymous source detection is studied in \cite{quiet-alerts, quiet-alerts-new}; however, these works require superconstant state space and only address this one task. 
Other research in wireless sensor networks investigates private data aggregation, which most closely resembles the goal of our research \cite{CMT2005, LLZC2013, GV2009}. As before, these works require high computation and local memory as they implement their solutions using homomorphic encryption. Where alternative methods are used to avoid relying on encryption, a specialized network topology is needed for success \cite{STE19} or only specific functions are computable \cite{GV2009}. 

While far from comprehensive, this sample of related works suggests that much of the research on privacy in wireless sensor networks is either limited by network topology or relies on computationally intensive encryption. For this reason, our goal is to develop privacy-preserving solutions for data aggregation in population protocols, bearing in mind the resource restrictions of the model.
\subsection{Contribution}

In this work, we study the privacy of population protocols in the random scheduling model. We demonstrate how existing privacy definitions fail under certain modelling assumptions, give new precise definitions of privacy in these settings, and offer a novel protocol in the uniform random scheduling population protocol model satisfying the new privacy definitions. In this work, we restrict our focus to computing the \Remainder\ predicate.
\section{Preliminaries} \label{sec:prelims}

A \term{population protocol} $\mathcal{P}$ is a tuple $(Q, \delta, \Sigma, \mathcal{I}, O, \mathcal{O})$ consisting of \term{state set} $Q$, \term{transition function} $\delta$, \term{input set} $\Sigma$, \term{input function} $\mathcal{I}$, \term{output set} $O$, and \term{output function} $\mathcal{O}$ \cite{AADFP2006}. Protocols are run by a population, which consists of a set of $n$ agents $\{A_j\}_{j=1}^n$ each with some input $i_j \in \Sigma$. 
At the start of the protocol, each agent converts its input to a state in $Q$ via $\mathcal{I}: \Sigma \rightarrow Q$. In the early population protocol literature, $\mathcal{I}$ is only ever considered to be a deterministic function; however, in this work, we extend the model to allow for $\mathcal{I}$ to be randomized.
The transition function $\delta: Q^2 \rightarrow Q^2$ designates how the agents update their states upon interacting with each other in pairs. As a shorthand for saying $\delta(q_1, q_2) = (q_1', q_2')$, we write $q_1, q_2 \rightarrow q_1', q_2'$ where $\delta$ is implied. The protocol aims to compute some function (whose output is in the output set $O$) on the initial inputs of the agents in the population. An agent's output value is a function of the agent's state, determined by $\mathcal{O}: Q \rightarrow O$.


The collection of agents' inputs is denoted as a vector $I \in \Sigma^n$, where each index of $I$ reflects the input of a particular agent in the population. Adopting terminology from \cite{DFGR2007}, we refer to $I$ as an \term{input vector}. When the size of the state space is $O(1)$, the protocol cannot distinguish between two agents in the same state nor with the same input; therefore, we may want to refer to the multiset of input values in the input vector $I$, denoted multiset$(I)$.
After converting these inputs to elements of $Q$, the global state of the population is called a \term{configuration} and is represented as a vector $C \in Q^n$, where the $i$-th entry of the vector denotes the state of the $i$-th agent. Abusing notation, we say that $\mathcal{I}(I)=\langle \mathcal{I}(i_j) \rangle_{j=1}^n$ is the configuration resulting from applying the input function $\mathcal{I}$ to each of the agent inputs in $I=\langle i_j \rangle_{j=1}^n$. 

Agents update their states via interactions with one another which are performed at discrete intervals, called \term{steps}. At each step, an ordered pair of agents $(A_i, A_j)$ is selected from the population by the \term{scheduler}. To distinguish between the two agents in the ordered pair, we call the first agent the \term{\Initiator} and the second the \term{\Responder}. 
When an interaction takes place, the two selected agents update their states according to the transition function $\delta$ which may change the counts of states in the population, thereby updating the configuration. Let $\mathcal{C}$ be the configuration space, or the set of all possible configurations for a population of $n$ agents with state space $Q$. We say that a configuration $D \in \mathcal{C}$ is \term{reachable} from $C \in \mathcal{C}$ via $\delta$ if there exists some series of ordered agent pairs such that starting from $C$, if the configuration is updated according to $\delta$ on those ordered pairs, then the resulting configuration is $D$ \cite{AADFP2006}. If $D$ is reachable from $C$, then we write $C \rightarrow D$.
The infinite sequence of configurations resulting from the scheduler's infinite choice of interaction pairs is called an \term{execution}. 
An execution of a protocol is said to \term{converge} at a step $\tau$ when, for every step $t > \tau$, the output of each agent's state at $t$ is the same as it is at $\tau$ (i.e. the output of every agent converges to some value and never changes thereafter). A stronger notion of termination is for a protocol to \term{stabilize}, meaning that after reaching some configuration $C^*$, the only configurations reachable from $C^*$ result in the same outputs at every agent as in $C^*$. Abusing notation, we say $\mathcal{O}(C) = \lambda$ (or, the output of the \textit{configuration} is $\lambda$) if $\mathcal{O}(q_j) = \lambda$ for every $q_j \in C$.

The goal of the population is to compute some function $\Phi$ on the input vector $I$, which means that the population eventually stabilizes towards a set of configurations $\mathcal{D} \subseteq \mathcal{C}$ for which $\mathcal{O}(D) = \Phi(I)$ for all $D \in \mathcal{D}$.
The results of our work are commensurable with those of \cite{DFGR2007} which demonstrate that the semilinear predicates, which can be expressed using \Threshold\ and \Remainder, can be computed with output independent privacy under adversarial scheduling. 
Our work focuses on \Remainder, defined for population protocols as follows:
\begin{definition}
Given positive integers $k$ and $n$, non-negative integer $r < k$, and input vector $I \in \mathbb{Z}_k^n$, let $\textnormal{\Remainder}(I) = \True$ iff $\sum_{j=1}^n i_j \equiv r \pmod k$.
\end{definition}

The scheduler determines the pair of agents that interact at each step. The scheduler's choice of agent pairs may either be adversarial or probabilistic.
An \term{adversarial scheduler} chooses pairs of agents to interact at each step as it desires, subject to a fairness condition. The condition used most commonly is called \term{strong global fairness}, and it states that if some configuration $C$ occurs infinitely often, and $C \rightarrow C'$, then $C'$ must occur infinitely often as well \cite{AADFP2006}. This means that if some configuration \textit{can} occur, it eventually \textit{must} occur, even if the adversarial scheduler wishes to delay its occurrence indefinitely. In works adopting adversarial scheduling, it can be claimed that a protocol eventually stabilizes to the correct answer, but not how quickly.
A random or \term{probabilistic scheduler} instead selects pairs of agents to interact with one another according to some fixed probability distribution (usually uniform) over the ordered pairs of agents. 
Although population protocols consider interactions to occur in sequence, the systems they model typically consist of agents participating in interactions in parallel. As such, a natural estimation of \term{parallel time} is to divide the total number of interactions by $n$, as this roughly estimates the expected number of interactions initiated by a particular agent in the population. Note that in population protocols with \textit{non-uniform} random scheduling, this notion of time is no longer necessarily suitable.


Our work crucially relies on distinguishing between an externally visible component of the agent state and a concealable secret state. Adopting notation from \cite{amir_et_al:LIPIcs:2020:13084}, we let $S$ be the \term{internal state} space and $M$ the set of \term{messages} which can be sent between the agents. Since each agent has both an internal and external state component, the total state space is then the Cartesian product of these sets $Q=S \times M$. This means that $\delta$ is instead a function computed locally at each agent according to its own state and the ``message received'' by its interacting partner $\delta: S \times M \times M \times \{\Initiator, \Responder\} \rightarrow S \times M$. This new mapping enforces the restriction that an agent can only use its received message to update its own state, and it does not observe the update to its interacting partner's state. For convenience, we use the original shorthand notation $\langle s_0, m_0 \rangle, \langle s_1, m_1 \rangle \rightarrow \langle s_0', m_0' \rangle, \langle s_1', m_1' \rangle$ to reflect the agents' state changes, where it is understood that the state update of $A_b$ is computed independently of $s_{1-b}$.
\section{Adversarial Model}
\label{sec:adversarial-models}

In order to evaluate the extent to which private information can be learned by an observer in a population protocol, we must define the nature of the observer and its capabilities. In this work, we consider the agent inputs to be private information. We will consider the observer to take the form of an agent interacting in the protocol, meaning that it can observe the population only as other agents do, i.e., by participating in interactions as they are slated by the scheduler. However, we do not preclude the possibility that the observer may have greater computational capabilities than ordinary honest agents.

We assume that the observer is \term{semi-honest}, meaning that it must adhere to the protocol rules exactly, but may try to infer additional knowledge from the system \cite{Lindell2017}. As such, the observer can only gather knowledge by interacting with other agents as prescribed by the transition function $\delta$.

Since an observer presents as an agent in the population, we can imagine that multiple adversaries may infiltrate the system. However, we restrict that each observer be \term{non-colluding}, meaning that it cannot communicate with other adversative nodes in the network besides participating in the protocol interactions honestly. This is because otherwise we could imagine that an observer may disguise itself as multiple agents in the population making up any fraction of the system. 
Although not studied within this work, it is of interest to find bounds on the fraction of agents that can be simulated by the observer in any network and still successfully hide honest agents' inputs.
Notice that the restriction that the observer is both semi-honest and non-colluding is equivalent to assuming that there is only a single adversative agent in the population, because from the point of view of the observer, all other agents appear to be honest.

Finally, we allow a distinction between externally visible messages and internally hidden states as in \cite{amir_et_al:LIPIcs:2020:13084} to allow agents to conceal a portion of their states toward the end goal of achieving privacy. The distinction between messages and the internal state will be crucial to studying privacy in the population model as without it, there is no mechanism for hiding information from an observer.
\section{Definitions of Input Privacy} \label{sec:privacy-defs}

In this section, we examine definitions of privacy in population protocols under adversarial and probabilistic scheduling given our specified adversarial model.



\subsection{Output Independent Privacy}
\label{subsec:old-def}

The privacy-preserving population protocol from \cite{DFGR2007} operates under the adversarial scheduling model and uses constant state-space. Therefore, \cite{DFGR2007} demonstrates privacy in the context of computing semilinear predicates only. The authors offer a formal definition of input privacy under these circumstances called \term{output independent privacy}, defined as follows:


\begin{displayquote}
    ``A population protocol has this property if and only if there is a constant $n_0$ such that for any agent $p$ and any inputs $I_1$ and $I_2$ of size at least $n_0$ in which $p$ has the same input, and any execution $E_1$ on input $I_1$, and any $T$, there exists an execution $E_2$ on input $I_2$, such that the histories of $p$’s interactions up to $T$ are identical in $E_1$ and $E_2$.''
\end{displayquote}
Essentially, this definition states that a semi-honest process $p$ cannot tell whether the input vector is $I_1$ or $I_2$ given its sequence of observations because either input could have yielded the same observations under an adversarial scheduler.

Output independent privacy is a successful measure in \cite{DFGR2007} because the scheduling in that work is assumed to be adversarial, therefore no inference can be made about the interaction pattern. The authors leverage this to achieve privacy which is best framed as ``plausible deniability'' -- an agent may directly observe another agent's input, but the unpredictability of the adversarial scheduler disallows the observer to claim with certainty that the observed value is indeed the input.

This argument breaks down when the scheduler is probabilistic because now an agent can infer a probability distribution on the interaction pattern, and thus also infer a probability distribution on the input value of the agent's interacting partner. In light of this insight, we now introduce novel definitions for the purpose of assessing privacy in population protocols with probabilistic scheduling.

\subsection{Definitions of Privacy Under Probabilistic Schedules} \label{subsec:new-def-privacy}


Consider an agent $A$ with initial state $q_0^A=(s_0^A, m_0^A)$. Given its sequence of observed messages and the role (\Initiator\ or \Responder) played by $A$ in each interaction, $A$ can deterministically compute each of its subsequent state updates. Let's call these messages (observed by $A$) $o^A_1, o^A_2, o^A_3, ...$, and denote by $q_\varepsilon^A = \delta(\rho_\varepsilon^A, s^A_{\varepsilon-1}, m^A_{\varepsilon-1}, o_\varepsilon^A) = (s^A_\varepsilon, m^A_\varepsilon)$ the updated state of $A$, originally in state $q_{\varepsilon-1}^A=(s^A_{\varepsilon-1}, m^A_{\varepsilon-1})$, upon interacting as $\rho_\varepsilon^A \in \{\Initiator, \Responder\}$ with another agent with message $o^A_\varepsilon$ in its $\varepsilon$-th interaction. 
Adopting notation from \cite{Lindell2017}, we denote the \term{view} of an agent $A$ participating in protocol $\mathcal{P}$ in an execution $E$ by \view$_A^\mathcal{P}(E)=\langle i_A; q_0^A; (\rho_1^A, o_1^A), (\rho_2^A, o_2^A), ... \rangle$. This view consists of $A$'s input, the initial state of $A$, and a list of $A$'s interactions over the course of the execution, from which every subsequent state of $A$ can be computed.\footnote{For randomized $\delta$, we assume $A$ has a fixed tape of random bits that it uses to update its state, so $A$ can still reconstruct its entire view from the specified information.}

Let $\bm{\bview_A^\mathcal{P}(C)}$ be a random variable representing the view of agent $A$ drawn uniformly from all realizable executions starting from configuration $C$ resulting from the possible randomness used by the scheduler. Similarly, let $\bm{\bview_A^\mathcal{P}(I)}$ be a random variable representing the view of agent $A$ drawn from all possible executions starting from any configuration $C$ in the range of $\mathcal{I}(I)$ according to the probability distribution given by the randomness of $\mathcal{I}$. In general, we use the convention that random variables appear in mathematical boldface.

Privacy, like many other security-related key terms, has a wide range of technical interpretations. As such, we now offer several distinct formal definitions of privacy in the population model.

\subsubsection{Plausible Deniability}

Perhaps the weakest form of privacy we can possibly define is that of \textit{plausible deniability}, meaning that an adversary always doubts its guess of an agent's input value (even if it has unbounded resources). This is not a novel concept \cite{DFGR2007, deniability}, but in the context of input vector privacy for probabilistic population protocols, we define this notion as follows:

Let $\mathcal{M}_{\lambda}=\{\multiset(I): \Phi(I)=\lambda\}$ be the set of all distinct multisets of inputs whose corresponding input vector evaluates to $\lambda$,\footnote{Recall that agents in the same state are indistinguishable by the protocol; therefore, $\Phi$ must map any input vectors with the same multiset of inputs to the same output.} and let $\mathcal{M}_\lambda^{\kappa}=\{\multiset(I) : \multiset(I) \in \mathcal{M}_{\lambda} \land \kappa \in \multiset(I)\}$ be the set of all distinct multisets of inputs outputting $\lambda$ which contain at least one input equal to $\kappa$.

\begin{definition} \label{def:weak-privacy}
Let $\mathcal{P}$ be a population protocol on $n$ agents with input set $\Sigma$ and let $\mathcal{D}$ be any probability distribution on input vectors in $\Sigma^n$. Then $\mathcal{P}$ is \term{weakly private} if for every distribution $\mathcal{D}$ on $\Sigma^n$, every non-colluding semi-honest unbounded agent $\A$ in a population of size $n$ executing $\mathcal{P}$, and for any view $V = \langle i ; q ; \{(\rho_\varepsilon^\A, o_\varepsilon^\A)\} \rangle$ with output $\lambda$ (as determined from the view $V$)
and with $\lvert \mathcal{M}_{\lambda}^{i} \rvert > 1$, there exist $I_1$ and $I_2$ in $\mathcal{S}_{\lambda}$ such that
\begin{enumerate}
    \item both $\multiset(I_1)$ and $\multiset(I_2)$ are elements of $\mathcal{M}_{\lambda}^{i}$,
    \item $\multiset(I_1) \neq \multiset(I_2)$, and
    \item $\Pr(\bview\bm{_\A^\mathcal{P}(I_1)}=V) = \Pr(\bm{\bview_\A^\mathcal{P}(I_2)}=V)$,
\end{enumerate}
where the probabilities in the final condition are taken over $\mathcal{D}$, the randomness of $\mathcal{I}$, and the uniform randomness of the scheduler.
\end{definition}

In plain English, Definition \ref{def:weak-privacy} says that any agent participating in the protocol cannot simply guess the ``most likely'' input vector because for each such vector, pending certain circumstances, there exists a distinct input vector yielding the same views for that agent with the same probabilities. This definition differs from output independent privacy \cite{DFGR2007} in that it considers adversarial strategies for guessing the input vector which rely on distributional data collected from interactions with other agents.

The condition $\lvert \mathcal{M}_{\lambda}^{i} \rvert > 1$ necessitates that weak privacy may only hold for multisets of inputs for which plausible deniability is even possible. For example, if the output of the computation for the \Or\ predicate is 0, then there is only one possible multiset of inputs that could have yielded this outcome, so there is no denying what the input vector must have been (namely, the all-zero vector).

\begin{toappendix}

\subsection{Alternative Definition of Privacy}

\subsubsection{Computational Indistinguishability}

This definition borrows concepts from classical cryptography dealing with proofs of zero-knowledge \cite{10.1145/22145.22178}, which rely on what is called the \term{simulation paradigm}. Essentially, the idea is that there are two versions of a protocol: A ``true'' version of the protocol is executed exactly as expected; however, an ``ideal'' version of the protocol is one where agents do not have inputs, but instead choose their inputs randomly based only on the output of the protocol. If an observer, say some agent in the population, cannot tell the difference between these two scenarios, then it must be the case that it learns nothing more about the inputs of other agents than it learns from information to which it already has access (namely, its own input and the output of the computation) because inputs in the latter scenario are chosen totally randomly subject to the protocol output remaining fixed.

Loosely speaking, ``indistinguishability'' in the population model means that a computationally bounded agent participating in the protocol does not learn more about any other agent's secret input than it does from its own input and the protocol output due to the fact that it cannot distinguish between a ``real'' execution of the protocol and an ``ideal'' one.

In order to define this formally, we first introduce some additional concepts:

\begin{definition}
A function $f$ is \term{negligible} \cite{katz-lindell} if for every polynomial $p(\cdot)$ there exists an $N$ such that for all $n > N$ it holds that $f(n)<p(n)$.
\end{definition}

In other words, we call $f$ negligible if it grows more slowly than every polynomial function.

\begin{definition}
Two distributions $\mathcal{X}$ and $\mathcal{Y}$ on a sample space $\Lambda$ of size $2^w$ are \term{computationally indistinguishable} \cite{katz-lindell} if for every probabilistic polynomial time (PPT) adversary $\A$ with oracle access to a polynomial (in $w$) number of instances from one of these two distributions, there exists a negligible function $\epsilon$ such that \[\lvert \Pr(\A^{\mathcal{X}(\cdot)}(w)=0) - \Pr(\A^{\mathcal{Y}(\cdot)}(w)=0) \rvert \leq \epsilon(w)\]
\end{definition}

That is to say, there does not exist a PPT algorithm which can reliably identify either of these two distributions when given access to only a polynomial number of samples from one of the distributions. If $\mathcal{X}$ and $\mathcal{Y}$ are computationally indistinguishable, then we write $\mathcal{X} \equiv \mathcal{Y}$.

Recall that an execution is an infinite sequence of configurations. Therefore, we denote the view of $A$ up until some fixed step $\tau$ as: \view$_A^\mathcal{P}(E)^{\tau}=\langle i_A; q_0^A; (\rho_1^A, o_1^A), (\rho_2^A, o_2^A), ..., (\rho_{\alpha_A}^A, o_{\alpha_A}^A) \rangle$, where $\alpha_A$ is the number of interactions $A$ participates in from the start of the execution to a fixed step $\tau$ of the execution.

Now, we define a notion of privacy related to computational indistinguishability in the population model:

\begin{definition}
A protocol $\mathcal{P}$ satisfies \term{input indistinguishability} if for every semi-honest non-colluding PPT agent $\A$ and any polynomial $\tau(n)$, there exists a simulator $S$ such that for every possible output $\lambda$ and every $I \in \mathcal{S}_{\lambda}$,
\[\{S(n, i_\A, \lambda)\} \equiv \{\bm{\bview_\A^\mathcal{P}(I)^{\tau(n)}} :  \Phi(I)=\lambda \land i_\A \in I\}\]
\end{definition}

In words, given only the computation output $\lambda$, a fixed agent $\A$ and their input $i_\A$, as well as the size of the population $n$, there exists a PPT simulator $S$ that can compute some polynomial-length prefix of a $\tau(n)$-length view at $\A$ that is computationally indistinguishable from $\bm{\bview_{\A}^\mathcal{P}(I)}$ in a real execution (up to some fixed step $\tau(n)$), so long as this view, computed by $S$, is consistent with inputs $n$, $i_\A$, and $\lambda$. By the simulation paradigm, this definition implies that no matter what the prefix of the execution, $\A$ learns as much about the input vector as it learns from its own input $i_\A$, public knowledge (such as $n$ and an approximation of time elapsed $\tau$), and the output of the computation $\lambda$.

Note that the definition claims that the property of indistinguishability holds true \textit{for each} input vector and is not just a property which holds true in the aggregate when averaged over the input vector space.

\subsection{Comparison of Definitions}

\end{toappendix}

\subsubsection{Information-Theoretic Input Privacy}

A stronger notion of privacy is one that claims that an observer cannot narrow down the possibility of input vectors at all based on its observations. This prompts our next definition.

Let $\mathcal{P}$ be a population protocol with input set $\Sigma$ and let $\mathcal{D}$ be a probability distribution on input vectors in $\Sigma^n$. Let $\bm{I} \sim \mathcal{D}$ be a random variable representing the selected input vector. Additionally, let $\bm{i_\A}$ and $\bm{\lambda_\A}$ be random variables representing the input and output at agent $\A$, and let $\bview\bm{_\A^\mathcal{P}(i, \lambda)}$ be a random variable representing the view of agent $\A$ participating in an honest execution of $\mathcal{P}$ that is consistent with a fixed input $i$ at $\A$ and observed output $\lambda$.

\begin{definition}
Protocol $\mathcal{P}$ satisfies \term{information-theoretic input privacy} if for every non-colluding semi-honest unbounded agent $\A$ and every input $i \in \Sigma$, output $\lambda \in O$, view $V$, input vector $I \in \mathcal{S}_{\lambda}$, and distribution $\mathcal{D}$ on $\Sigma^n$,
\[Pr(\bm{I} = I \mid \bview\bm{_\A^\mathcal{P}(i, \lambda)}=V) = Pr(\bm{I} = I \mid \bm{i_\A}=i, \bm{\lambda_\A}=\lambda),\]
where $V$ is consistent with input $i$ and output $\lambda$.
\end{definition}

The above definition essentially states that conditioned on knowing one's own input and the output of the computation, the rest of the agent's view in the protocol's computation gives no advantage in guessing the input vector.

\emptysec We offer an alternative definition of privacy that is independent of our main results called \term{input indistinguishability} that can be found in the Appendix.



Intuitively, it is straightforward to see that information-theoretic privacy is the strongest of the definitions discussed in this section (see Appendix for proof):

\begin{theoremrep} \label{theorem:it-privacy-is-strongest}
If $\mathcal{P}$ is information-theoretically private, then $\mathcal{P}$ also satisfies output independent privacy, weak privacy, and input indistinguishability.
\end{theoremrep}

\begin{proof}
We prove this claim in three parts, in each one by showing the contrapositive.

\begin{enumerate}[label=\roman*.]
    \item \textit{If $\mathcal{P}$ does not satisfy output independent privacy, then it is not information-theoretically private.}

    Because $\mathcal{P}$ does not satisfy output independent privacy, there do not exist any two input vectors and corresponding executions that yield the same views at a process $p$. For the sake of contradiction, suppose that $\mathcal{P}$ is weakly private. Then there must exist input vectors $I_1$ and $I_2$ such that \[\Pr(\bview\bm{_\A^\mathcal{P}(I_1)}=V) = \Pr(\bview\bm{_\A^\mathcal{P}(I_2)}=V)\]
    However, this would contradict the assumption that $\mathcal{P}$ is not output independent private. Therefore, $\mathcal{P}$ must not be weakly private, and by (ii), it must also not be information-theoretically private.
    \item \textit{If $\mathcal{P}$ is not weakly private, then it is not information-theoretically private.}
    
    Because $\mathcal{P}$ is not weakly private, there exists some adversary $\A$ with view $V=\langle i ; q ; \{(\rho_\varepsilon^\A, o_\varepsilon^\A)\} \rangle$ and output $\lambda$ such that $\lvert \mathcal{M}_\lambda^{i} \rvert > 1$, but there is no pair of input vectors $I_1,I_2 \in \mathcal{S}_{\lambda}^2$ such that all three conditions in Definition \ref{def:weak-privacy} hold. Thus for every pair of input vectors $I_1$ and $I_2$ with distinct multisets that are members of $\mathcal{M}_{\lambda}^{i}$ (i.e. they are consistent with the adversary's input $i$ and the computation output $\lambda$), we have that
    \[\Pr(\bview\bm{_\A^\mathcal{P}(I_1)}=V) \neq \Pr(\bview\bm{_\A^\mathcal{P}(I_2)}=V)\]
    As a result, we can say that
    \begin{align*}
        \Pr(\bm{I}=I_1 \mid \bview\bm{_\A^\mathcal{P}(i, \lambda)} = V) &= \frac{\Pr(\bm{I}=I_1 \land \bview\bm{_\A^\mathcal{P}(i, \lambda)} = V)}{\Pr(\bview\bm{_\A^\mathcal{P}(i, \lambda)} = V)} \\
        &= \Pr(\bview\bm{_\A^\mathcal{P}(I_1)}=V)/\Pr(\bview\bm{_\A^\mathcal{P}(i, \lambda)} = V) \\
        &\neq \Pr(\bview\bm{_\A^\mathcal{P}(I_2)}=V)/\Pr(\bview\bm{_\A^\mathcal{P}(i, \lambda)} = V) \\
        &= \Pr(\bm{I}=I_2 \mid \bview\bm{_\A^\mathcal{P}(i, \lambda)} = V)
    \end{align*}
    Using this information, we will show that for some distribution $\mathcal{D}$ on $\Sigma^n$ it is the case that \[Pr(\bm{I} = I \mid \bview\bm{_\A^\mathcal{P}(i, \lambda)}=V) \neq Pr(\bm{I} = I \mid \bm{i_\A}=i, \bm{\lambda_\A}=\lambda)\] for some $I\in \Sigma^n$, indicating that $\mathcal{P}$ is not information-theoretically private.

    Consider the uniform distribution on $\Sigma^n$, which means that for all $I \in \Sigma^n$ $\Pr(\bm{I}=I)=1/\lvert \Sigma \rvert ^n$, so for any pair of input vectors $I_1, I_2$ in $\mathcal{M}_{\lambda}^{i}$, \[\Pr(\bm{I}=I_1)=\Pr(\bm{I}=I_2)\] Due to the fact that these inputs come from $\mathcal{M}_{\lambda}^{i}$, we can also say that \[\Pr(\bm{I}=I_1 \mid \bm{i_\A}=i, \bm{\lambda_\A}=\lambda)=\Pr(\bm{I}=I_2 \mid \bm{i_\A}=i, \bm{\lambda_\A}=\lambda)\] However, because \[\Pr(\bm{I}=I_1 \mid \bview\bm{_\A^\mathcal{P}(i, \lambda)} = V) \neq \Pr(\bm{I}=I_2 \mid \bview\bm{_\A^\mathcal{P}(i, \lambda)} = V)\] for some $I_1, I_2$ in $\mathcal{M}_{\lambda}^{i} \subseteq \mathcal{S}_{\lambda}$, we must have that either 
    \[Pr(\bm{I} = I_1 \mid \bview\bm{_\A^\mathcal{P}(i, \lambda)}=V) \neq Pr(\bm{I} = I_1 \mid \bm{i_\A}=i, \bm{\lambda_\A}=\lambda)\]
    or
    \[Pr(\bm{I} = I_2 \mid \bview\bm{_\A^\mathcal{P}(i, \lambda)}=V) \neq Pr(\bm{I} = I_2 \mid \bm{i_\A}=i, \bm{\lambda_\A}=\lambda)\]
    Otherwise, we would have to have
    \begin{align*}
        \Pr(\bm{I}=I_1 \mid \bview\bm{_\A^\mathcal{P}(i, \lambda)}=V)&=\Pr(\bm{I}=I_1 \mid \bm{i_\A}=i, \bm{\lambda_\A}=\lambda)\\
        &=\Pr(\bm{I}=I_2 \mid \bm{i_\A}=i, \bm{\lambda_\A}=\lambda)\\
        &=\Pr(\bm{I}=I_2 \mid \bview\bm{_\A^\mathcal{P}(i, \lambda)}=V)
    \end{align*}
    which would contradict our initial assumption that $\mathcal{P}$ is not weakly private. Thus, there exists some $I \in \mathcal{S}_{\lambda}$ and distribution on $\Sigma^n$ for which \[\Pr(\bm{I} = I \mid \bview\bm{_\A^\mathcal{P}(i, \lambda)}=V) \neq Pr(\bm{I} = I \mid \bm{i_\A}=i, \bm{\lambda_\A}=\lambda)\] so $\mathcal{P}$ must also not be information-theoretically private.
    \item \textit{If $\mathcal{P}$ is not input indistinguishable, then it is not information-theoretically private.}
    
    To make these two definitions comparable, let's assume that the adversary always has unbounded computational power. Even if this is the case, we show that information-theoretic privacy is still stronger than unbounded input indistinguishability (and so weakening the adversary as in the definition of computational indistinguishability yields the same relationship between these two definitions).
    
    Let $\mathcal{V}=\Sigma \times Q \times \{\{\Initiator, \Responder\} \times M\}^{\tau(n)}$ (for some fixed $\tau$ that is polynomial in $n$) be the space of all possible views of an agent in an execution of $\mathcal{P}$ with $\tau(n)$ interactions in a view. Denote by $\mathcal{X}(I)$ the distribution $\{S(n, i, \lambda)\}$ over $\mathcal{V}$; and denote by $\mathcal{Y}(I)$ the distribution $\{\bview\bm{_\A^\mathcal{P}(I)^{\tau(n)}} :  \Phi(I)=\lambda \land i \in I\}$ over $\mathcal{V}$. By assumption, $\mathcal{P}$ is not input indistinguishable, so there exists some input vector $I^*$ for which $\mathcal{X}(I^*) \not\equiv \mathcal{Y}(I^*)$, as in $\Pr(\mathcal{X}(I^*)=V)\neq \Pr(\mathcal{Y}(I^*)=V)$ for some $V \in \mathcal{V}$. Suppose this $I^*$ has input $i$ at $\A$ and output $\lambda$.
    
    Because $\mathcal{Y}$ reflects the real distribution of views for an agent $\A$ with input $i$ and output $\lambda$ when $I^*$ is the true input vector, we have
    \[\Pr(\mathcal{Y}=V)=\Pr(\bview\bm{_\A^\mathcal{P}(i, \lambda)}=V \land \bm{I}=I^*)\]
    In addition, because $\A$ is computationally unbounded, it can simulate the protocol on every possible input vector using every possible random schedule (for at least $\tau(n)$ observations at $\A$) to compute the exact probability with which input vectors containing $i$ and outputting $\lambda$ yield view $V$, so
    \[\Pr(\mathcal{X}=V)=\Pr(\bview\bm{_\A^\mathcal{P}(i, \lambda)}=V \land \bm{i_\A}=i, \bm{\lambda_\A}=\lambda)\]
    Further, since $i \in I^*$ and $\Phi(I^*)=\lambda$, we know that \[\Pr(\bview\bm{_\A^\mathcal{P}(i, \lambda)}=V \mid \bm{I}=I^*) \leq \Pr(\bview\bm{_\A^\mathcal{P}(i, \lambda)}=V \mid \bm{i_\A}=i, \bm{\lambda_\A}=\lambda)\] However, this must actually be a strict inequality because otherwise $I^*$ would be the only input vector in $\mathcal{M}_\lambda^{i}$ which would imply \[\Pr(\bm{i_\A}=i, \bm{\lambda_\A}=\lambda \land \bview\bm{_\A^\mathcal{P}(i, \lambda)}=V)=\Pr(\bm{I}=I^* \land \bview\bm{_\A^\mathcal{P}(i, \lambda)}=V)\] and therefore also $\Pr(\mathcal{X}=V)=\Pr(\mathcal{Y}=V)$ (contradicting our initial assumption).

    Therefore, (taking $\varkappa$ to be the event that $\bm{i_\A}=i$ and $\bm{\lambda_\A}=\lambda$) we have
    \begin{align*}
        &\ \Pr(\bm{I}=I^* \mid \bview\bm{_\A^\mathcal{P}(i, \lambda)}=V) \\
        =&\ \frac{\Pr(\bview\bm{_\A^\mathcal{P}(i, \lambda)}=V \mid \bm{I}=I^*)\Pr(\bm{I}=I^*)}{\Pr(\bview\bm{_\A^\mathcal{P}(i, \lambda)}=V)}\\
        =&\ \frac{\Pr(\bview\bm{_\A^\mathcal{P}(i, \lambda)}=V \mid \bm{I}=I^*)\Pr(\bm{I}=I^* \land \bm{i_\A}=i, \bm{\lambda_\A}=\lambda)}{\Pr(\bview\bm{_\A^\mathcal{P}(i, \lambda)}=V)}\\
        =&\ \frac{\Pr(\bview\bm{_\A^\mathcal{P}(i, \lambda)}=V \mid \bm{I}=I^*)\Pr(\bm{I}=I^* \land \varkappa)}{\Pr(\bview\bm{_\A^\mathcal{P}(i, \lambda)}=V)}\\
        =&\ \frac{\Pr(\bview\bm{_\A^\mathcal{P}(i, \lambda)}=V \mid \bm{I}=I^*)\Pr(\bm{I}=I^* \mid \varkappa)\Pr(\varkappa)}{\Pr(\bview\bm{_\A^\mathcal{P}(i, \lambda)}=V)}\\
        <&\ \frac{\Pr(\bview\bm{_\A^\mathcal{P}(i, \lambda)}=V \mid \varkappa)\Pr(\bm{I}=I^* \mid \varkappa)\Pr(\varkappa)}{\Pr(\bview\bm{_\A^\mathcal{P}(i, \lambda)}=V)}\\
        =&\ \frac{\Pr(\bview\bm{_\A^\mathcal{P}(i, \lambda)}=V \land \varkappa)\Pr(\bm{I}=I^* \mid \varkappa)}{\Pr(\bview\bm{_\A^\mathcal{P}(i, \lambda)}=V)}\\
        \leq&\ \frac{\Pr(\bview\bm{_\A^\mathcal{P}(i, \lambda)}=V)\Pr(\bm{I}=I^* \mid \varkappa)}{\Pr(\bview\bm{_\A^\mathcal{P}(i, \lambda)}=V)}\\
        =&\ \Pr(\bm{I}=I^* \mid \varkappa)\\
        =&\ \Pr(\bm{I}=I^* \mid \bm{i_\A}=i, \bm{\lambda_\A}=\lambda)
    \end{align*}
    Thus we have shown that $\mathcal{P}$ must also not be information-theoretically private.
\end{enumerate}
\end{proof}

\section{Private \Remainder\ with Adversarial Scheduling}

As a means for comparison, we analyze the \Remainder\ protocol from \cite{DFGR2007}, shown in Algorithm~\ref{fig:dfgr07remainder}. 
The protocol does not distinguish between internal state space and message space, so the entirety of each agent's state is seen by its interacting partner. 
The agent states are tuples $(v, f)$, where $v$ is the value of the agent and $f$ is a flag bit denoting whether or not the agent has decided its output yet. The protocol accumulates the total sum (modulo $k$) of all agents' inputs by transferring values in units rather than in full in a single interaction. As shown in (M1), the protocol subtracts 1 from one of the inputs and adds it to the other input, maintaining the invariant that the sum of all the values in the population is the same at each step. Because all computations are done modulo $k$, (M1) can be repeated indefinitely. Transitions (M2) and (M3) handle the flag bit, ensuring that (M1) occurs an unbounded but finite number of times.

\begin{table}[t]
    \centering
    \renewcommand{\arraystretch}{1.6}
    \begin{tabular}{c}\hline
    \textbf{Algorithm 1: Output Independent Private \Remainder\ \cite{DFGR2007}}\\\hline\\[-15pt]
    {$\!\begin{aligned}
        (v_1, 1), (v_2, 1) &\rightarrow (v_1+1, 1), (v_2-1, 1) &\text{(M1)}\\[1pt]
        (*, 1), (*, *) &\rightarrow (*,0), (*,*) &\text{(M2)}\\[1pt]
        (*,0), (*,1) &\rightarrow (*,1), (*,1) &\text{(M3)}\\[1pt]
        (v_1,0), (v_2,0) &\rightarrow (v_1+v_2,0), (0,0) &\text{(M4)}\\[1pt]
        (v_1,0), (0,0) &\rightarrow (v_1,0), (\bot_0, 0) &\text{(M5)}\\[1pt]
        (\bot_i, *), (*,1) &\rightarrow (0,0), (*,1) &\text{(M6)}\\[1pt]
        (r,0), (\bot_i,0) &\rightarrow (r,0), (\bot_1, 0) &\text{(M7)}\\[1pt]
        (v_1,0), (\bot_i,0) &\rightarrow (v_1,0), (\bot_0,0)\text{, if }v_1\neq r &\text{(M8)}\\[1pt]
    \end{aligned}$}
    \\\hline
    \end{tabular}
    \captionsetup{labelformat=empty}
    \caption{Each agent $A_j$ has initial state $(i_j, 1)$. Arithmetic is done modulo $k$, and $*$ is a wildcard that can match any value. The output values are $\{\bot_0, \bot_1\}$, denoting that the predicate is \textsc{False} or \textsc{True}, respectively. The protocol converges when all but one agent has $\bot_0$ or $\bot_1$ as their state.}
    \label{fig:dfgr07remainder}
\end{table}


The crux of the proof that Algorithm~\ref{fig:dfgr07remainder} satisfies output independent privacy focuses on transition (M1). When an adversarial process $p$ interacts with an honest agent $A$ in state $(v, f)$, $p$ cannot know how close $v$ is to $A$'s original input because, for $n \geq 3$, we can construct multiple executions wherein $A$ has value $v$ upon interacting with $p$. For example, we can construct an execution where some agent $B$ transfers as many units to $A$ via (M1) as needed to get $A$'s value to be $v$, and as long as $p$ and $B$ do not interact with each other before $p$ interacts with $A$, $p$'s view is the same in this execution. 


However, output independent privacy does not successfully carry over to the random scheduling model because we can no longer construct \textit{any} execution ``fooling'' the process $p$, as some such executions are of very low probability. For instance, the probability that agents $A$ and $B$ interact $v'$ times in a row, during which time $p$ does not interact with $B$ at all, becomes small for large values of $v'$. This means that it is less probable that an agent's value will deviate from its original input value early on in the execution. A formal proof of the algorithm's lack of privacy is given in the Appendix.

\begin{toappendix}

\begin{theorem} \label{theorem:secretive-birds-not-private}
Algorithm~\ref{fig:dfgr07remainder} is not information-theoretically input private under a uniform random scheduler.
\end{theorem}

\begin{proof}
First, we observe that an agent can compute the probability with which it interacts with another agent in their initial state under the uniform random scheduler conditioned on the event that it is also that agent's first interaction. Fix an agent $A_j$ and consider its first interaction. With probability $2/n$, agent $A_j$ participates in an interaction at any particular step. Conditioned on this event at step $t$, let us compute the probability that this is the first interaction for \textit{both} agents in the execution (meaning that neither agent has been selected to interact at any prior step). Under the uniform scheduler, this probability is equal to
\[\left(\frac{\text{\# pairs not including the two agents}}{\text{\# total possible pairs}}\right)^{t-1}=\left(\frac{\binom{n-2}{2}}{\binom{n}{2}}\right)^{t-1}=\left(\frac{n^2-5n+6}{n^2-n}\right)^{t-1}\]
Summing this over all possible steps at which an interaction with $A_j$ can occur, we have that the probability that $A_j$'s first interaction is with another agent which has also never interacted before (and is therefore still in their initial state) is \[\sum_{t=1}^{\infty} \frac{2}{n}\left(\frac{n^2-5n+6}{n^2-n}\right)^{t-1} = \frac{n-1}{2n-3} > \frac{1}{2}\]

Thus, with probability at least 1/2, an agent can view another agent's initial state. Although an agent cannot necessarily detect that this is the case, the fact that this happens with such a large probability changes the conditional probability that an observed input exists in the population conditioned on the view of an agent:

The probability that any particular agent sampled uniformly from the population has input $\sigma$ is
\begin{align*}
    p_{\sigma}=\frac{1}{n}\sum_{j=1}^n \Pr(\bm{i_j} = \sigma) 
    &= \frac{1}{n}\sum_{I \in \Sigma^n} \left( \sum_{j=1}^n \Pr(\bm{i_j} = \sigma \mid \bm{I}=I)\right) \Pr(\bm{I}=I) \\
    &= \frac{1}{n}\sum_{I \in \Sigma^n} (\text{count of }\sigma\text{ in }I) \Pr(\bm{I}=I)
\end{align*}
Given that $\sum_{\sigma \in \Sigma}p_{\sigma}=1$, there must be some $\sigma\in\Sigma$ for which $p_{\sigma} \leq 1/\lvert \Sigma \rvert$, and for any non-trivial population protocol $\lvert\Sigma\rvert \geq 2$. Thus, given the strategy described above, the probability that an agent's first observed state is the actual input of its interacting partner is strictly greater than 1/2, and for some observed input in the input set, this is not equal to its \textit{a priori} probability. As such, Algorithm~\ref{fig:dfgr07remainder} is not information-theoretically private.
\end{proof}

The protocols for solving \Threshold\ and \Or\ from \cite{DFGR2007} can similarly be proven to lack privacy. We omit these proofs for concision.

\subsubsection{Note on Input Indistinguishability and Weak Privacy}

Notice that the Algorithm~\ref{fig:dfgr07remainder} protocol does not attribute probabilities to the nondeterministic transitions in $\delta$. For example, two agents in states $(0, 1)$ and $(1, 1)$ could take transition (M1) or transition (M2) from Algorithm~\ref{fig:dfgr07remainder}; however, we have no probabilistic distribution from which to draw this decision. Therefore, it makes it more complicated to ascertain whether or not this algorithm satisfies these definitions of input privacy.

The notion of privacy presented in \cite{DFGR2007} relies on there being an unbounded number of times that (M1) could be applied to any particular agent's state, thereby transforming the input vector without restriction. If the probability of transition (M1) is too small, then agents' state values (while of the form $(*, 1)$) are confined to a random walk near their bona fide input values. Conversely, even if the probability of transition (M1) is large enough, the first few interactions still reveal the input values of each agent because the entirety of an agent's state is viewed by its interacting partner with non-negligible probability, as demonstrated previously in the evaluation of information-theoretic privacy. In either case, the view of an agent narrows the input vector space significantly, sometimes down to a single multiset of inputs.

For example, consider the case where $k=2$. Then any view of $\A$ resulting from an execution wherein two agents besides $\A$ interact with one another can be obfuscated by supposing that the order of the agents in the interaction is flipped. However, in executions where the only such interactions involve one agent in state 0 and another agent in state 1, this actually does not yield a new multiset of inputs (but rather just swaps the inputs held by the two agents). In fact, the only way to obfuscate the input multiset would be to change both of the inputs of two agents with equal inputs, but this would necessarily have to change the view of $\A$. While this case is contrived and rare, it still precludes Algorithm~\ref{fig:dfgr07remainder} from being even weakly private. This motivates the need for novel protocols satisfying these privacy definitions in the probabilistic scheduling population model.
\end{toappendix}
\section{Private \Remainder\ with Probabilistic Scheduling}

In this section, we introduce a novel algorithm for information-theoretically privately computing \Remainder\ in the population protocol model with probabilistic scheduling.
Our algorithm is inspired by the famous example of cryptographically secure multiparty computation of \Remainder\ in a ring network. We refer to this algorithm as \protocol{RingRemainder}, and it works as follows:


There are $n$ agents $A_1,...,A_n$ arranged in a circle. Agent $A_1$ performs the leader's role, which is to add a uniformly random element $r \in \mathbb{Z}_k$ to their input and pass the sum (modulo $k$) to agent $A_2$. For each remaining agent $A_i$, upon receiving a value from $A_{i-1}$, $A_i$ adds its own input to that value and passes the resulting sum to $A_{i+1 \pmod n}$. When $A_1$ receives a value from $A_n$, it subtracts $r$ and broadcasts the result to everyone. Suppose the agents have inputs $i_1, ..., i_n$. Then $A_1$ sends $m_1 = i_1 + r$ to $A_2$, $A_2$ sends $m_2 = i_1 + r + i_2$ to $A_3$, and so on, until $A_n$ sends $m_n = r + \sum_{j=1}^n i_j$ to $A_1$. Thus, the value broadcast to all agents $m_n-r$ is exactly equal to $\sum_{j=1}^n i_j$, the sum of the agents' inputs modulo $k$.
Assuming honest participants and secure pairwise communication,
this protocol achieves information-theoretic input privacy (see Appendix for proof).

\begin{toappendix}
    \subsection{\protocol{RingRemainder}}

\begin{theorem}
    \protocol{RingRemainder} is information-theoretically private.
\end{theorem}

\begin{proof}
Let $\bm{I}=\langle \bm{i_1}, ..., \bm{i_n} \rangle$ be a random variable representing the input vector to the protocol and let $\bm{m_j}$ be a random variable representing the message in $\mathbb{Z}_k$ passed from agent $A_j$ to $A_{j+1}$.

Assuming secure peer-to-peer communication and non-colluding participants, each agent $A_j$ only learns its own input, its received message, and the final value broadcast by the leader (which exactly equals the output of the computation). Thus the view of $A_j$ consists of three values in $\mathbb{Z}_k$.

We know $\bm{m_{j}} = \bm{r} + \sum_{x=1}^{j} i_x \pmod k$, where $\bm{r}$ is a random variable representing the uniformly drawn group element of $\mathbb{Z}_k$. Thus, 
\[\Pr\left(\bm{m_{j}} = m^* \mid \bm{I}=I\right) = \Pr\left(\bm{r} = m^*-\sum_{x=1}^{j} i_x \pmod k\right)=1/k\]
and \[\Pr(\bm{m_{j}}=m^*) = \sum_{I \in \Sigma^n} \Pr(\bm{m_{j}}=m^* \mid \bm{I}=I) \Pr(\bm{I}=I) = 1/k \sum_{I \in \Sigma^n} \Pr(\bm{I}=I) = 1/k\]
As such, we may conclude that $\bm{I}$ and $\bm{m_{j}}$ are independent (and this holds true for every $j \in \{1, ..., n\}$).

Consider the view $V = \langle i^*, m^*, \lambda^* \rangle$ at $A_j$. Then
\begin{align*}
    \Pr(\bm{I}=I \mid \bview\bm{_\A^{\mathcal{P}}(i_j, \lambda_j)}=V) &= \Pr(\bm{I}=I \mid \bm{i_j}=i^*, \bm{m_{j-1}}=m^*, \bm{\lambda_j}=\lambda^*) \\
    &= \Pr(\bm{I}=I \mid \bm{i_j}=i^*, \bm{\lambda_j}=\lambda^*)
\end{align*}
because $\bm{I}$ and $\bm{m_{j-1}}$ are independent. As a result, we have \[\Pr(\bm{I}=I \mid \bm{i_j}=i^*, \bm{\lambda_j}=\lambda^*)=\Pr(\bm{I}=I \mid \bview\bm{_\A^{\mathcal{P}}(i_j, \lambda)}=V)\] for every $V$ consistent with input $i^*$ and output $\lambda^*$ (regardless of the value of $m^*$), as needed.
In short, learning any agent's input (or any partial sum along the ring) reduces to guessing the exact random element added to the aggregate sum by $A_1$.
\end{proof}
\end{toappendix}

We now adapt this scheme to compute \Remainder\ in the population model with information-theoretic privacy.

\subsubsection{Algorithm Overview}

Our protocol simulates the transfer of information exactly as in \protocol{RingRemainder}. We assume that the protocol has an initial leader with a special token that circulates the population. Each time an agent receives the token and some accompanying value, it adds its input to that value and passes the sum, along with the token, to another agent. This means the current owner of the token holds the aggregate sum of the agents' inputs who previously held the token. When an agent passes the token to another agent, it labels itself as ``visited'' so as to ensure that its input is included in the sum exactly one time. Once the token has visited all of the agents, it is returned to the leader (along with the total sum of all of the agents' inputs). In order to achieve this functionality, there are two crucial obstacles we must overcome:

First, we need a mechanism for securely transferring a message between two agents such that no other agent learns the message except the sender and the intended recipient. This task is nontrivial because population protocols do not allow agents to verify a condition before transmitting a message in an interaction; it is assumed that the message exchange and state update occur instantaneously. 
To do this, we provide a secure peer-to-peer transfer subroutine in Section~\ref{subsec:p2p-transfer}.

Second, we need a way to determine whether or not every agent in the population has been visited by the token. When this happens, we want the final token owner to pass the token back to the leader so that the leader can remove the randomness it initially added to the aggregate that has been passed among the agents. We must try to prevent passing the aggregate back to the leader before all inputs have been incorporated into the aggregate as this would cause some agents to be excluded from the computation. 
In order to achieve this, we use the probing protocol from \cite{AAE2006} which we describe in further detail in Section~\ref{subsec:probe}.

Leveraging these two subroutines, we design our main algorithm for computing \Remainder\ with information-theoretic privacy in Section~\ref{subsec:remainder}.

\subsection{Secure Peer-to-Peer Transfer} \label{subsec:p2p-transfer}

In order for our algorithm to guarantee input privacy, the communication of the intermediate sums between any two agents must remain secure. Here we introduce a novel secure peer-to-peer transfer protocol, defined as follows:

\begin{definition}
Let $M$ be a message space, $\mathcal{D}$ be some distribution on $M$, and $I$ be any fixed input vector in $\Sigma^n$. A \term{secure peer-to-peer transfer routine} is a protocol $\mathcal{P}$ that transfers data $m \xleftarrow{\mathcal{D}} M$ from one agent $\Sender$ to another $\Receiver$ such that there exist PPT algorithms $W_1, W_2$ where
\[\Pr\left(W_1(\bview\bm{^\mathcal{P}_\Sender(I)})=m\right)=\Pr\left(W_2(\bview\bm{^\mathcal{P}_\Receiver(I)})=m\right)=1\]
and for all $i: A_i\not\in \{\Sender, \Receiver\}$ and PPT algorithm $W'$
\[\Pr\left(W'(\bview\bm{^\mathcal{P}_{A_i}(I)})=m\right)=\Pr(m \xleftarrow{\mathcal{D}} M)\]
\end{definition}
In other words, a secure peer-to-peer transfer routine allows a \Sender\ to transfer a message $m$ to a \Receiver\ such that only \Sender\ and \Receiver\ are privy to $m$ and all other agents cannot guess $m$ with any advantage over knowing only the \textit{a priori} distribution on the message space.

Our Algorithm~\ref{fig:p2p-transfer} satisfies this definition: Each agent's state $\langle \mu, (r, L) \rangle$ consists of a hidden secret $\mu$, and a public randomness value $r$ and label $L$. The goal of the protocol is to pass a secret message from one agent (marked as \Sender\ with label $\mf{S}$, of which there may only be one in the population) to another agent meeting some specified criteria labeled by $\mf{u}$, of which there may be any number (including zero).
Until the \Sender\ meets an agent with label $\mf{u}$, it refreshes its randomness at each interaction to ensure that the randomness it transmits to the \Receiver\ is uniformly random (S1). When the \Sender\ finally meets some agent with $\mf{u}$, it marks that agent as the \Receiver\ and transmits its fresh randomness value $r$; it also updates its own token to $\mf{S}'$ to remember that it has met and labeled a \Receiver\ (S2). Then, the \Sender\ waits to meet the \Receiver\ again, at which point it gives it a message masked with the randomness it sent in the previous interaction and marks itself with the label $\omf{u}$ to signify the end of the transmission (S3). By the end of the protocol, exactly one agent is selected as the \Receiver\ and stores $\mu$ internally. The protocol has state space $(\mathbb{Z}_k \cup \{\bot\})^2 \times \{\mf{S}, \mf{S}', \mf{R}, \mf{u}, \omf{u}\}$, which for constant $k$ is of size $O(1)$. We prove that Algorithm~\ref{fig:p2p-transfer} is a secure peer-to-peer transfer routine in the Appendix:

\begin{table}[t]
    \centering
    \renewcommand{\arraystretch}{1.6}
    \begin{tabular}{c}\hline
    \textbf{Algorithm 2: Population Protocol for Secure P2P Transfer}\\\hline\\[-15pt]
    {$\!\begin{aligned}
        \langle \mu , (r, \mf{S}) \rangle, \langle * , (*, \omf{u}) \rangle &\rightarrow \langle \mu , (r', \mf{S}) \rangle, \langle * , (*, \omf{u}) \rangle &\text{(S1)}\\[1pt]
        \langle \mu , (r, \mf{S}) \rangle, \langle * , (*, \mf{u}) \rangle &\rightarrow \langle \bot , (\mu - r, \mf{S}') \rangle, \langle r , (*, \mf{R}) \rangle &\text{(S2)} \\[1pt]
        \langle \bot , (x, \mf{S}') \rangle, \langle y, (*, \mf{R}) \rangle &\rightarrow \langle \bot , (\bot, \omf{u}) \rangle, \langle x + y, (*, \mf{S}) \rangle &\text{(S3)}\\[1pt]
    \end{aligned}$}
    \\\hline
    \end{tabular}
    \captionsetup{labelformat=empty}
    \caption{Secure peer-to-peer transfer protocol.}
    \label{fig:p2p-transfer}
\end{table}


\begin{toappendix}
    \subsection{Secure Peer-to-peer Transfer}
\end{toappendix}

\begin{theoremrep} \label{theorem:secure-p2p}
Algorithm~\ref{fig:p2p-transfer} is a secure peer-to-peer transfer routine.
\end{theoremrep}

\begin{proof}
First note that the label component $L$ of the message state is updated independently of $\mu$ and therefore leaks no information about $\mu$. Thus we will focus our attention to the state updates of the hidden secret and the public randomness.

There is one specified \Sender\ with token $\mf{S}$ and some number of eligible \Receiver\ agents. By (S1), the randomness used in the \Sender's external message is refreshed for each interaction, so when an agent is finally labeled as the designated \Receiver\ in (S2), the random value which it copies to its internal state is known (with probability 1) only by the \Sender\ and the \Receiver. Let $\bm{r} \leftarrow \mathbb{Z}_k$ be a random variable representing the actual random value transferred in this transition between the \Sender\ and the \Receiver.

At this point, the \Sender\ changes its external message to be $\bm{s}=\mu-\bm{r} \pmod k$. Recall that $\bm{r}$ is uniformly drawn from $\mathbb{Z}_k$, so for all $x \in \mathbb{Z}_k$ it is the case that $\Pr(\bm{r}=x)=1/k$. As such, for all $x \in \mathbb{Z}_k$, $\Pr(\bm{s} = \mu-\bm{r} = x)=\Pr(\bm{r} = \mu-x)=1/k$, so $\bm{s}$ is also uniformly distributed over $\mathbb{Z}_k$ (regardless of the a priori distribution $\mathcal{D}$ from which the hidden secret $\mu$ is drawn). Because the \Receiver\ is the only other agent that knows the exact value of $\bm{r}$, it learns $\mu$ on its next interaction with the \Sender. 

Define $W_1$ to be an algorithm that selects the hidden secret of a \Sender\ agent when it interacts as the \Initiator\ with another agent with label $\mf{u}$, and define $W_2$ to be an algorithm that selects the hidden secret of a \Receiver\ agent when it interacts as the \Responder\ with another agent with label $\mf{S}'$. Then, \[\Pr\left(W_1(\bview\bm{^\mathcal{P}_\Sender(I)})=\mu\right)=\Pr\left(W_2(\bview\bm{^\mathcal{P}_\Receiver(I)})=\mu\right)=1\]

However, for every other agent $A_j$ in the population, every observed external message (including $\mu-\bm{r}$) appears to be a uniformly random value, independent of the distribution from which the hidden secret $\mu$ is drawn. Thus any $W'$ must guess the value of $\mu$ in the absence of additional information beyond the \textit{a priori} distribution on the message space: \[\Pr\left(W'(\bview\bm{^\mathcal{P}_{A_j}(I)})=\mu\right)=\Pr(\mu \xleftarrow{\mathcal{D}} \mathbb{Z}_k)\]
As a result, only the \Sender\ and the chosen \Receiver\ can know $\mu$, making Algorithm~\ref{fig:p2p-transfer} a secure peer-to-peer transfer routine.
\end{proof}

\subsection{Probing Protocol} \label{subsec:probe}

In order to adapt \protocol{RingRemainder} to the population protocol model, we need a way to detect when every agent has been included in the aggregation so the final sum can be passed back to the leader. To do this, we use a probe.

\begin{toappendix}

\subsection{Probing Protocol}

\begin{table}[t]
    \centering
    \renewcommand{\arraystretch}{1.6}
    \begin{tabular}{c}\hline
    \textbf{Algorithm 4: Population Protocol for Epidemic-Driven Probing}\\\hline\\[-15pt]
    {$\!\begin{aligned}
        x, y &\rightarrow x, \max(x, y) & \text{(P1)}\\[1pt]
        0, y &\rightarrow 0, y & \text{(P2)}\\[1pt]
        x, y &\rightarrow x, 2 \quad [x > 0] & \text{(P3)}\\[1pt]
    \end{aligned}$}
    \\\hline
    \end{tabular}
    \captionsetup{labelformat=empty}
    \caption{Transition rules for the probing protocol from \cite{AAE2006}, where $x$ and $y$ are in $\{0,1,2\}$.}
    \label{fig:probe}
\end{table}

Algorithm~\ref{fig:probe} gives transition rules for performing a probe for a predicate $\pi$ using three states, 0, 1, and 2: When the \Responder\ does not satisfy $\pi$, the agents transition via (P1) and when the \Responder\ does satisfy $\pi$, they transition according to (P2) and (P3). The following lemma gives bounds on the running time of the probing protocol in Algorithm~\ref{fig:probe} and quantifies its probability of error:

\begin{lemma}[see \cite{AAE2006}, Lemma 6] \label{lemma:aae06-lem6}
For any $c > 0$, there is a constant $d$ such that for sufficiently large $n$, with probability at least $1-n^{-c}$ it is the case that after $dn\ln n$ interactions in the probing protocol either (a) no agent satisfies the predicate and every agent is
in state 1, or (b) some agent satisfies the predicate and every
agent is in state 2.
\end{lemma}

\subsection*{}

\end{toappendix}

A \term{probing protocol}, or \term{probe}, is a population protocol that detects the existence of an agent in the population satisfying a given predicate \cite{AAE2006}. In essence, the probe (initiated by the leader) sends out a 1-signal through a population of agents in state 0. If the 1-signal reaches an agent satisfying the predicate, that agent initiates a 2-signal which spreads back to the leader by epidemic. Higher number epidemics overwrite lower ones, so if some agent in the population satisfies $\pi$ then the leader eventually sees the 2-signal. The probe, used in conjunction with the phase clock from the same work \cite{AAE2006}, allows the leader to detect the presence of an agent satisfying $\pi$ in $O(n \log n)$ interactions using $O(1)$ states with probability $1-n^{-c}$ for any fixed constant $c>0$ (see Appendix).

We define the ``output'' of the protocol (computed only at the leader) to be 0 for states 0 and 1, and 1 for state 2 (i.e. the leader's probe outputs 1 if and only if some agent in the population satisfies $\pi$). At the start of each round of the phase clock, agents reset their value to 0 and the leader initiates a new probe. Both the probe and the phase clock states are components of the message space, and the transitions for these subroutines are independent of the transitions for the main protocol, so we consider the two ``protocols'' to be taking place in parallel.

\subsection{\Remainder\ with Information-Theoretic Privacy} \label{subsec:remainder}

We provide here a novel algorithm which computes \Remainder\ and achieves \textit{information-theoretic input privacy} in the population protocol model with high probability, assuming a uniform random scheduler. 

First, each agent applies the input function $\mathcal{I}$ to their input as follows:
\[\mathcal{I}(i_j, \ell) = \begin{cases}
    \langle i_j + r^0, (r^j, \mf{S}, 1, Z=Z_0) \rangle &\ell=1\\
    \langle i_j, (r^j, \mf{u}, 0, Z=Z_0) \rangle &\ell=0\\
\end{cases}\]
where $r^j$ is drawn uniformly at random from $\mathbb{Z}_k$ for $j \in \{0, 1, ..., n\}$, and $Z$ (initialized to $Z_0$) is a probe subroutine (including its associate phase clock). The input function assumes an initial leader, specified by $\ell=1$. The components of the state $\langle \mu, (r, L, \ell, Z) \rangle$ are $\mu$ (the hidden internal component of the state called the \term{secret}), $r$ (the \term{mask}), $L$ (the agent's \term{label}), $\ell$ (the \term{leader bit}), and $Z$ (the \term{probe}). The transitions describing the protocol can be found in Algorithm~\ref{fig:private-remainder}.

The general structure of the transitions from the secure peer-to-peer transfer protocol in Algorithm~\ref{fig:p2p-transfer} is used to send the intermediate sums in (R1), (R2), and (R3). However, instead of just storing the message received, the \Receiver\ computes the sum of the message and its own input and stores the result internally. Each subsequent \Sender\ searches the population for an agent whose input has not yet been incorporated into the sum (signified by the $\mf{u}$ state). When no one in the population has $\mf{u}$ anymore, the probe detects this and outputs 1 at the leader from this point onward.

When the probe begins to output 1, with high probability every agents' label is set to $\omf{u}$, alerting the leader to set its label to $\mf{u}$. This makes the leader the only agent able to be the next \Receiver. When the leader receives the final value stored at the \Sender, the leader can place the answer into a separate portion of the external state (not shown in Algorithm~\ref{fig:private-remainder}) so that all other agents can copy it, which takes $O(n^2 \log n)$ additional steps with high probability. The leader must also have an additional component to its \textit{hidden} state which stores the randomness used in its initial message transfer (also not shown in Algorithm~\ref{fig:private-remainder}).

\begin{table}[t]
    \centering
    \renewcommand{\arraystretch}{1.6}
    \begin{tabular}{c}\hline
    \textbf{Algorithm 3: Information-Theoretically Private \Remainder}\\\hline\\[-15pt]
    {$\!\begin{aligned}
        \langle *, (r, \mf{S}, *, *) \rangle, \langle *, (*, \omf{u}, *, *) \rangle &\rightarrow \langle *, (r', \mf{S}, *, *) \rangle, \langle *, (*, \omf{u}, *, *) \rangle &\text{(R1)}\\[1pt]
        \langle u, (r, \mf{S}, *, *) \rangle, \langle v, (*, \mf{u}, *, *) \rangle &\rightarrow \langle \bot, (u - r, \mf{S}', *, *) \rangle, \langle v + r, (*, \mf{R}, *, *) \rangle &\text{(R2)}\\[1pt]
        \langle *, (x, \mf{S}', *, *) \rangle, \langle y, (*, \mf{R}, *, *) \rangle &\rightarrow \langle *, (\bot, \omf{u}, *, *) \rangle, \langle x + y, (*, \mf{S}, *, *) \rangle &\text{(R3)}\\[1pt]
        \langle \bot, (\bot, \omf{u}, 1, 1) \rangle, \langle *, (*, *, *, *) \rangle &\rightarrow \langle \bot, (\bot, \mf{u}, 1, 1) \rangle, \langle *, (*, *, *, *) \rangle &\text{(R4)}\\[1pt]
    \end{aligned}$}
    \\\hline
    \end{tabular}
    \captionsetup{labelformat=empty}
    \caption{Although not shown, each interaction also performs an update to the probing subroutine by advancing the phase clock and probe at both agents in the interaction. Transitions (R1-R3) are analogous to transitions (S1-S3) from Algorithm~\ref{fig:p2p-transfer}.}
    \label{fig:private-remainder}
\end{table}

\begin{toappendix}

\subsection{Algorithm~\ref{fig:private-remainder} Correctness Proof}

From Algorithm~\ref{fig:private-remainder}, we can see that the effective steps of the computation are driven by the \Sender\ via transitions (R1), (R2), and (R3). The only transition not involving the \Sender\ is (R4), wherein the leader sets its label from $\omf{u}$ to $\mf{u}$ when it detects that every other agent has label $\omf{u}$, signifying that every agent's value is aggregated into a single sum held by the current \Sender\ (this only happens once). When such a point is reached, that \Sender\ should now pass that sum back to the leader via (R2) and (R3), i.e. through secure peer-to-peer transfer. We aim to prove that Algorithm~\ref{fig:private-remainder} computes \Remainder\ in $\Theta(n^3 \log n)$ steps with high probability. First, we will prove a sequence of lemmas that we will later use to prove the overall correctness and running time of Algorithm~\ref{fig:private-remainder} in Theorem~\ref{theorem:prob-rem-correct}.

\begin{lemma} \label{lemma:uniq-S}
In any configuration of a population executing Algorithm~\ref{fig:private-remainder}, exactly one agent has label $\mf{S}$ or $\mf{S}'$.
\end{lemma}

\begin{proof}
In the initial configuration, this is true by construction of the input function $\mathcal{I}$. At any later step, if we assume that exactly one agent in the population has $\mf{S}$ or $\mf{S}'$, then it either:
\begin{itemize}
    \item maintains ownership of the unique token $\mf{S}$ via (R1),
    \item updates its token to $\mf{S}'$ (thereby eliminating the only $\mf{S}$ in the population and replacing it with $\mf{S}'$) via (R2), or 
    \item transfers this token (originally as $\mf{S}'$) to another agent (now $\mf{S}$) via (R3).
\end{itemize}
Transition (R4) does not involve agents with the label $\mf{S}$ nor $\mf{S}'$. Thus by induction, this claim is true at every step of any execution of Algorithm~\ref{fig:private-remainder}.
\end{proof}

\begin{lemma} \label{lemma:trans-r2-time}
For any $c>0$, starting from a configuration $C$ with exactly one agent having label $\mf{S}$, exactly $t$ agents with label $\mf{u}$, and the remaining agents with label $\omf{u}$, with probability at least $1-n^{-c}$, these labels remain unchanged for $\Theta(\frac{c n^2 \log n}{t})$ steps, at which point transition (R2) is performed.
\end{lemma}

\begin{proof}
The probability that the scheduler selects an \Initiator\ with label $\mf{S}$ and a \Responder\ with label $\mf{u}$ (resulting in transition (R2)) is $\frac{1}{n} \cdot \frac{t}{n-1}$. Let $\bm{X}$ be a random variable representing the number of steps for (R2) to occur from this starting configuration. By independence and uniform randomness of the scheduler, $\bm{X}$ is geometric with parameter $p=\frac{t}{n(n-1)}$, so $\bm{X}=\frac{n(n-1)}{t}$ in expectation, or is at most $k$ with probability
\begin{align*}
    \Pr(\bm{X} \leq k) = \sum_{j=0}^k \left(1-p\right)^j p
    &= \sum_{j=0}^k \left(1-\frac{t}{n(n-1)}\right)^j \frac{t}{n(n-1)}\\
    &= 1-\left(1-\frac{t}{n(n-1)}\right)^{k+1}
\end{align*}
For $k=\frac{n(n-1) \cdot c \log n}{t}-1$, this probability is at least $1-e^{-c \log n}$ (as a result of the fact that $(1-1/x)^x < 1/e$ for $x \geq 1$).

The only other non-null transition that can possibly occur from $C$ is between labels $\mf{S}$ and $\mf{u}$ (thereby eliciting transition (R1)), which has no effect on the labels of the agents in the population. Therefore, the labels in the population do not change until (R2) is executed once, which occurs in $\Theta(cn^2 \log n / t)$ steps with probability $1-e^{-c \log n}=1-n^{-c}$ for any choice of $c>0$.
\end{proof}

\begin{lemma} \label{lemma:trans-r3-time}
For any $c>0$, starting from a configuration with exactly one agent having label $\mf{S}'$ and exactly one agent having label $\mf{R}$, we execute transition (R3) in $\Theta(cn^2 \log n)$ steps with probability $1-n^{-c}$.
\end{lemma}

\begin{proof}
By Lemma~\ref{lemma:uniq-S}, this agent with label $\mf{S}'$ is the only agent having either label $\mf{S}$ or $\mf{S}'$ in the population. Therefore, transition (R3) can eventually be taken when the agents with $\mf{S}'$ and $\mf{R}$ meet. Any particular pair of agents interacts with probability $1/\binom{n}{2}$. By the uniform randomness of the probabilistic scheduler, we expect to wait $O(n^2)$ steps on average for this event to occur. Moreover, the probability that this event occurs within $cn^2 \log n$ steps is at least $1-e^{-c \log n}=1-n^{-c}$ for any choice of $c>0$.
\end{proof}

Due to the fact that (R2) results in a configuration with exactly one $\mf{S}'$ and one $\mf{R}$, we have the following corollary:

\begin{corollary} \label{cor:r2-then-r3}
For any $c>0$, after any step where (R2) occurs, (R3) must also occur within $\Theta(cn^2 \log n)$ steps with probability $1-n^{-c}$.
\end{corollary}

\begin{lemma} \label{lemma:rem-val-transfer}
Transitions (R2) and (R3) aggregate the sum of the secrets of the \Sender\ and the \Receiver\ modulo $k$ into the secret of the \Receiver\ in $\Theta(cn^2 \log n)$ steps with probability $1-n^{-c}$ for any $c>0$.
\end{lemma}

\begin{proof}
By Corollary~\ref{cor:r2-then-r3}, (R2) and (R3) are essentially executed as a pair (taking $\Theta(cn^2 \log n)$ steps with high probability). Let the step where (R2) is executed between two agents $A_i$ (as the \Initiator) and $A_j$ (as the \Responder) be $\tau_0$. The secrets of these agents are $u$ and $v$ respectively, thus we aim to show that after (R2) and (R3) are performed, $A_j$ has secret $u+v \pmod k$.

Agent $A_i$ has label $\mf{S}$ and is therefore the \Sender, and upon executing transition (R2) we have that $A_j$ is labeled as the \Receiver. After this transition, we also have that the \Sender\ relabels itself with $\mf{S}'$ to signify that a \Receiver\ has been selected. $A_i$ has secret $\bot$ and mask $u+r$, and $A_j$ has secret $v-r$.

By Lemma~\ref{lemma:trans-r3-time} and Corollary~\ref{cor:r2-then-r3}, the next state update involving these two agents executes (R3) and this occurs within $\Theta(cn^2 \log n)$ steps with probability $1-n^{-c}$ for any $c>0$. The only other effective step which can occur is (R4) because every other transition involves some agent with a $\mf{S}$ label and by Lemma~\ref{lemma:uniq-S} there is no other $\mf{S}$ label in the population. Moreover, (R4) does not change the state of $A_i$ nor $A_j$ because their labels are not equal to $\omf{u}$. 

We know that when (R3) is eventually performed, this must be with $A_i$ as the \Initiator\ and $A_j$ as the \Responder\ because these are the only two agents with the appropriate labels in the population and no other effective steps have been taken to transfer them to any other agents. Thus, $A_i$ still has $u+r$ as its mask and $A_j$ has $v-r$ as its secret. Transition (R3) then results in $A_j$ updating its secret to $x+y=(u-r)+(v+r)=u+v$ and leaves $A_i$ with label $\omf{u}$ (indicating that its secret is now $\bot$).
\end{proof}

\begin{lemma} \label{lemma:inductive-step}
For any $c>0$, starting from a configuration $C$ with exactly one agent having label $\mf{S}$, exactly $t$ agents with label $\mf{u}$, and the remaining agents with label $\omf{u}$, with probability at least $1-n^{-c}$, we reach a configuration $C'$ with exactly one agent having label $\mf{S}$, exactly $t-1$ agents with label $\mf{u}$, and the remaining agents with label $\omf{u}$ in $\Theta(c n^2 \log n)$ steps.
\end{lemma}

\begin{proof}
By Lemma~\ref{lemma:trans-r2-time}, for any $c'>0$ and configuration $C$, we execute (R2) in $\Theta(c'n^2 \log n/t)$ steps with probability $1-n^{-c'}$ (only changing the randomness values for some number of agents). When transition (R2) is executed, the labels $\mf{S}$ and $\mf{u}$ are overwritten by $\mf{S}'$ and $\mf{R}$. By Lemma~\ref{lemma:trans-r3-time}, for any $c''>0$ we execute (R3) in $\Theta(c''n^2 \log n)$ additional steps from this configuration with probability $1-n^{-c''}$, at which point the labels $\mf{S}'$ and $\mf{R}$ are overwritten by $\omf{u}$ and $\mf{S}$. All in all, the net change in these $\Theta(c'n^2 \log n /t+c''n^2 \log n)$ steps is a decrease in the number of agents with label $\mf{u}$ by 1 (resulting in an increase in the number of agents with label $\omf{u}$ by 1). For any $c>0$, let $c'=c''=c+1$. Taking the union bound over the probability of failure for each phase described in these lemmas gives a total probability of failure of $2n^{-(c+1)}$ which is at most $n^{-c}$ for $c>0$ and $n \geq 2$. Plugging this expression for $c'$ and $c''$ in terms of $c$ back into the above asymptotic bound on steps, and knowing $1 \leq 1+1/t \leq 2$ for $t>0$, we have a total run time of \[\Theta(c'n^2 \log n /t+c''n^2 \log n) = \Theta((c+1)(1+1/t)n^2 \log n) = \Theta(cn^2 \log n)\]
\end{proof}

\begin{lemma} \label{lemma:agg-sum}
For any $c>0$, in $\Theta(cn^3 \log n)$ steps, with probability $1-n^{-c}$ Algorithm~\ref{fig:private-remainder} aggregates all agent secrets into one sum modulo $k$ and labels all agents with $\omf{u}$ except one agent marked as the \Sender\ holding the sum.
\end{lemma}

\begin{proof}
At the start of the protocol, each agent $A_j$ has state $\langle i_j, (r^j, \mf{u}, 0, 0) \rangle$ except the leader who has state $\langle i_j+r^0, (r^j, \mf{S}, 1, 0) \rangle$. This means that exactly one agent has label $\mf{S}$ while $n-1$ agents have label $\mf{u}$.

The only transition that can be taken is (R2) followed by (R3), which aggregates the sum of the leader's secret with some other agent in the population by Lemma~\ref{lemma:rem-val-transfer}. This also results in the leader being labeled with $\omf{u}$ and this selected agent being the new \Sender\ (reducing the number of agents with $\mf{u}$ by 1).

By Lemma~\ref{lemma:inductive-step}, for any $c'>0$, it takes $\Theta(c'n^2 \log n)$ steps to reduce the number of agents with $\mf{u}$ by 1 with probability $1-n^{-c'}$. By applying Lemma~\ref{lemma:inductive-step} repeatedly, we find that it takes
\[\Theta \left(\sum_{t=1}^{n-1} c' n^2 \log n  \right) = \Theta(c'(n-1)n^2 \log n) = \Theta(c'n^3 \log n) \]
steps for the number of agents with label $\mf{u}$ to become 0. Each time the number of agents with label $\mf{u}$ decrements, the agent whose label is overwritten has its secret aggregated into the secret of the \Sender\ and is then labeled with $\omf{u}$ to mark itself so it does not become a \Sender\ again. When no agent has $\mf{u}$, the \Sender\ has the total sum of the agent secrets. Taking a union bound over the probability of failure each time we applied Lemma~\ref{lemma:inductive-step}, we have that the total probability of failure is $n \cdot n^{-c'}$. For any $c>0$, letting $c'=c+1$ yields the desired result for $n \geq 1$.
\end{proof}

\begin{lemma} \label{lemma:zero-detector}
Transition (R4) only occurs once, when the probe determines that no agent in the population has label $\mf{u}$. For any $c>0$, with probability $1-n^{-c}$, the probe only goes off when all agents have $\omf{u}$.
\end{lemma}

\begin{proof}
By construction, there is only one leader in the population and every agent is initialized with a probe that outputs 0 upon instantiation. Once the probe goes off, it cannot be reset. Only when the probe goes off can (R4) occur, so this can only happen once and only to the leader.

By Lemma~\ref{lemma:aae06-lem6}, when some agent in the population satisfies $\pi$, for any $c'>0$, only with probability $n^{-c'}$ do we have that not every agent is in state 2. If one such agent is the leader who is probing for the label $\mf{u}$, then it mistakenly outputs 0 at the end of the phase clock cycle and causes the protocol to skip some number of agents in the protocol computation which may result in an incorrect output. In order to avoid this outcome, we need for the leader's probe to succeed for every phase clock round over the course of the protocol. By Lemma~\ref{lemma:agg-sum}, for any $c''>0$ it takes $\Theta(c''n^3 \log n)$ steps to aggregate all agent inputs into one sum with probability $1-n^{-c''}$. For $c''>0$, because a phase clock has $O(n \log n)$ steps in a round, we need the probe to succeed $O(c''n^2)$ consecutive times. By the union bound, the probability that the probe fails at least once in these $O(c''n^2)$ consecutive trials is $dc''n^2 \cdot n^{-c'}$ for some constant $d$. Thus for any $c>0$, letting $c'=c+3$ gives the desired result for sufficiently large $n$.
\end{proof}

Now, we are ready to prove the following theorem regarding the correctness of Algorithm~\ref{fig:private-remainder}:
\end{toappendix}

The correctness of Algorithm~\ref{fig:private-remainder} is stated below and proven in the Appendix:
\begin{theoremrep} \label{theorem:prob-rem-correct}
For any fixed $c>0$, Algorithm~\ref{fig:private-remainder} computes \textnormal{\Remainder} in a population of size $n$ in $\Theta(n^3 \log n)$ steps with probability at least $1-n^{-c}$.
\end{theoremrep}

\begin{proof}
By Lemma~\ref{lemma:agg-sum}, for any $c'>0$, in $\Theta(c'n^3 \log n)$ steps (with probability $1-n^{-c'}$), the population aggregates all agent secrets into one sum held by a single agent with the singular \Sender\ label, at which point all other agents are marked with $\omf{u}$ and the \Sender\ has secret
\[r^0 + \sum_{j=1}^n i_j \pmod k\]
By Lemma~\ref{lemma:aae06-lem6}, for any $c''>0$, the probe does not go off before this point and successfully does go off within $O(c''n \log n)$ steps of reaching this point with probability $1-n^{-c''}$. Therefore, by Lemma~\ref{lemma:zero-detector}, for any $c'''>0$ (R4) is then executed (which takes constant time in expectation and polylogarithmic time with probability $1-n^{-c'''}$), changing only the leader's label to $\mf{u}$. This allows the one remaining agent with the $\mf{S}$ token to perform one last effective exchange with the leader, which, by Lemma~\ref{lemma:rem-val-transfer}, transfers the aggregate sum back to the leader in $\Theta(c''''n^2 \log n)$ additional steps with probability $1-n^{-c''''}$ for any $c''''>0$.

Taking the union bound over these four sources of error gives us a probability of failure of $n^{-c'}+n^{-c''}+n^{-c'''}-n^{-c''''}$. For any fixed $c>0$, letting $c'=c''=c'''=c''''=c+1$ gives a probability of success that is at least $1-n^{-c}$ for sufficiently large $n$.

Finally, the leader can remove $r^0$ from the secret to obtain the remainder of the sum of agents' inputs modulo $k$. A separate mechanism may be used to distribute this value to the rest of the population by epidemic.
\end{proof}

Finally, we summarize the privacy guarantee of Algorithm~\ref{fig:private-remainder} in our final theorem and defer the proof to the Appendix:

\begin{toappendix}
\subsection{Algorithm~\ref{fig:private-remainder} Privacy Proof}
\end{toappendix}

\begin{theoremrep} \label{theorem:rem-info-theoretic-private}
When Algorithm~\ref{fig:private-remainder} correctly computes the \Remainder\ predicate, it satisfies information-theoretic input privacy.
\end{theoremrep}

\begin{proof}
Assuming that the protocol correctly computes the \Remainder\ predicate (i.e. the phase clock does not experience any failures until at least the time $T$ past convergence of the protocol), an agent's view consists of its own input, its initial state, and some number of observed interactions (from which all other information to which it has access, such as the protocol output, can be derived).

Let us consider the view of an agent in the population $A_j$ until some step $\tau > T$, where the agent has input $(i^*, \ell^*)$ and output $\lambda^*$, letting $\alpha_j(\tau)$ be the number of interactions in which $A_j$ participates until step $\tau$ (so that $\sum_{j=1}^n \alpha_j(\tau) = \tau$):
\[\bview\bm{_{A_j}^{\mathcal{P}}((i^*, \ell^*), \lambda^*)^{\tau}} = \langle (i^*, \ell^*) ; \bm{q_0^j} ; \{\bm{(\rho_t, (r_t, L_t, \ell_t, Z_t))}\}_{t=1}^{\alpha_j(\tau)} \rangle\]
Recall that $\mathcal{I}$ is randomized, so $\bm{q_0^j}$ is a random variable representing the actual initial state of agent $A_j$ as determined by the randomness used to compute $\mathcal{I}$.

Denote by $\mathcal{Z}$ the state space of the probing protocol, and let \[\mathcal{V}=\mathbb{Z}_k \times \{0,1\} \times \mathbb{Z}_k \times \{\{\Initiator, \Responder\} \times \mathbb{Z}_k \times \{\mf{S}, \mf{S}', \mf{u}, \omf{u}\} \times \{0,1\} \times \mathcal{Z}\}^{\alpha_j(\tau)}\] be the space of all possible views of $A_j$ until step $\tau$ of the computation.

Additionally, let $\bm{\theta_t} = \bm{(\rho_t, (r_t, L_t, \ell_t, Z_t))}$ be a random variable representing the $t$-th observation of $A_j$. Among the sequence $\{\bm{\theta_t}\}_{t=1}^{\alpha_j(\tau)}$, there are four special observations that must appear in any correct execution of the protocol:
\begin{align*}
    \theta_a &= (\Initiator, (r_a, \mf{S}, \ell_a, Z_a)) \text{, \Sender\ initiates secure transfer}\\
    \theta_b &= (\Initiator, (r_b, \mf{S}', \ell_b, Z_b))\text{, \Sender\ transmits secret value}\\
    \theta_c &= (\Responder, (r_c, \mf{u}, \ell_c, Z_c))\text{, eligible \Receiver\ is selected}\\
    \theta_d &= (\Responder, (r_d, \mf{R}, \ell_d, Z_d))\text{, secret sent to selected \Receiver}
\end{align*}
These observations occur in the above order for everyone except the leader (who instead views: $\theta_c, \theta_d, \theta_a, \theta_b$). The remaining observations either cause the \Sender\ to refresh its randomness value, the leader to relabel itself with $\omf{u}$, or do not result in a state change.

We want to show that the vector of values $\langle \bm{q_0^j}, \{\bm{\theta_t}\}_{t=1}^{\alpha_j(\tau)} \rangle$ is conditionally independent of the input vector $\bm{I}$ given the fixed input value $(i^*, \ell^*)$ at $A_j$ and the output $\lambda^*$ observed by $A_j$, which would imply
\begin{align*}
    &\ \Pr(\bm{I}=I^* \mid \bview\bm{_{A_j}^{\mathcal{P}}((i^*, \ell^*), \lambda^*)^{\tau}}=V) \\
    =&\ \Pr(\bm{I}=I^* \mid \bm{i_j} = (i^*, \ell^*), \langle \bm{q_0^j}, \{\bm{\theta_t}\}_{t=1}^{\alpha_j(\tau)} \rangle = \langle q^*, \{\theta_t^*\}_{t=1}^{\alpha_j(\tau)} \rangle, \bm{\lambda_j}=\lambda^*) \\
    =&\ \Pr(\bm{I}=I^* \mid \bm{i_j} = (i^*, \ell^*), \bm{\lambda_j}=\lambda^*)
\end{align*}
In order to show this, we first restrict our attention to a fixed schedule $s \in S$ (i.e. a specific choice of agent pairs selected to interact for $\tau$ steps of the protocol), where \[S = \{\{(x_t, y_t)\}_{t=1}^\tau : (x_t, y_t) \in \mathbb{Z}_n^2 \land x_t \neq y_t\}\] Let us assume that we only consider schedules $s$ wherein the protocol succeeds (i.e. all but some polynomially small fraction of the schedules). 

Because $\bm{q_0^j}$ is a function of $(i^*, \ell^*)$, we can say that the event $\bm{i_j}=(i^*, \ell^*) \land \bm{q_0^j}=q^*=\langle i^*, (r^*, \_, \ell^*, Z_0) \rangle$ is exactly the same as the event $\bm{i_j}=(i^*, \ell^*) \land \bm{r^j}=r^*$, where $\bm{r^j}$ is a random variable representing the initial randomness used by $A_j$ in its computation of $\mathcal{I}$. Recall that $\bm{r^j}$ is uniform over $\mathbb{Z}_k$ and independent of all other variables in the protocol (including $\bm{I}, \bm{i_j}, \bm{\lambda_j}$, and $s$). 

Furthermore, when a schedule $s$ is fixed, the order of agent pairs selected to interact determines the order in which the \Sender\ and \Receiver\ tokens are passed among the agents. This fully determines the $\bm{L_t}, \bm{\ell_t},$ and $\bm{Z_t}$ components of the $\bm{\theta_t}$. The only remaining values to consider are the external randomness value $\bm{r_t}$ and the internal secret state $\bm{\mu_t}$ (which is correlated with the value of $\bm{r_t}$ due to (R2)). However, by the assumption that the protocol successfully computes the \Remainder, we can see that (R2) is only executed once at each agent when the agent observes $\theta_a$ followed sometime afterwards by $\theta_b$ (which invoke transitions (R2) and (R3), respectively). For fixed $s$, the number of interactions between these two observations is fixed, and therefore the only correlation between these two values is that they are separated by the secret being transferred. For a fixed schedule (where the order in which agent values are aggregated is fixed), recall that this secret is equal to $r^0+\sum_{x \in X \subseteq \mathbb{Z}_n} i_x$. All other observations have $\bm{r_t}$ that is drawn uniformly and independently of all other values. Therefore, the event $\{\bm{\theta_t}\}_{t=1}^{\alpha_j(\tau)} \land s$ is the same as $\bm{r^0}=r^{**} \land \{\bm{r_t}=r_t^*\}_{t=1}^{\alpha_j(\tau)} \land s$, where $r^{**}=r_t^*-\sum_{x \in X \subseteq \mathbb{Z}_n} i_x \pmod k$ and each $r_t^*$ is the exact observed randomness value in each observation made by $A_j$. This is a consequence of the fact that given a fixed schedule, the only variation in the values of each of the $\bm{\theta_t}$ is from the randomness values drawn by the agents at each step, which are also uniformly random and independent of all other variables of the protocol.

Let $\bm{S}$ be a random variable representing the $\tau$ ordered pairs of agents chosen to interact by the scheduler. Conditioned on some fixed schedule $s$ and some fixed view $V$ at $A_j$, and by the mutual independence of the uniformly drawn randomness values with all other variables in the protocol, for any $I \in \Sigma^n$ that is consistent with input $(i^*, \ell^*)$ at $A_j$ and output $\lambda^*$,
\begin{align*}
    &\ \Pr(\bm{I}=I \mid \bview\bm{_{A_j}^{\mathcal{P}}((i^*, \ell^*), \lambda^*)^{\tau}}=V, \bm{S}=s) \\
    =&\ \Pr(\bm{I}=I \mid \bm{i_j}=(i^*, \ell^*), \langle \bm{q_0^j}, \{\bm{\theta_t}\}_{t=1}^{\alpha_j(\tau)}\rangle=\langle q^*, \{\theta_t^*\}_{t=1}^{\alpha_j(\tau)}\rangle, \bm{\lambda_j}=\lambda^*, \bm{S}=s) \\
    =&\ \Pr(\bm{I}=I \mid \bm{i_j}=(i^*, \ell^*), \bm{r^j}=r^*, \{\bm{r_t}=r_t^*\}_{t=1}^{\alpha_j(\tau)}, \bm{\lambda_j}=\lambda^*, \bm{S}=s)\\
    =&\ \Pr(\bm{I}=I \mid \bm{i_j}=(i^*, \ell^*), \bm{\lambda_j}=\lambda^*, \bm{S}=s)
\end{align*}
Thus $\bm{I}$ is conditionally independent of $\langle \bm{q_0^j}, \{\bm{\theta_t}\}_{t=1}^{\alpha_j(\tau)} \rangle$ given $\bm{S}$, $\bm{i_j}$, and $\bm{\lambda_j}$. By the law of total probability, summing over all possible schedules $s \in S$ gives
\begin{align*}
    &\ \Pr(\bm{I}=I \mid \bview\bm{_{A_j}^{\mathcal{P}}((i^*, \ell^*), \lambda^*)^{\tau}}=V) \\
    =&\ \sum_{s \in S} \Pr(\bm{I}=I \mid \bview\bm{_{A_j}^{\mathcal{P}}((i^*, \ell^*), \lambda^*)^{\tau}}=V, \bm{S}=s)\Pr(\bm{S}=s) \\
    =&\ \sum_{s \in S} \Pr(\bm{I}=I \mid \bm{i_j}=(i^*, \ell^*), \bm{\lambda_j}=\lambda^*, \bm{S}=s)\Pr(\bm{S}=s) \\
    =&\ \sum_{s \in S} \Pr(\bm{I}=I \mid \bm{i_j}=(i^*, \ell^*), \bm{\lambda_j}=\lambda^*)\Pr(\bm{S}=s) \text{, by independence of the scheduler} \\
    =&\ \Pr(\bm{I}=I \mid \bm{i_j}=(i^*, \ell^*), \bm{\lambda_j}=\lambda^*) \sum_{s \in S} \Pr(\bm{S}=s)\\
    =&\ \Pr(\bm{I}=I \mid \bm{i_j}=(i^*, \ell^*), \bm{\lambda_j}=\lambda^*)
\end{align*}
\end{proof}

If the protocol fails due to a phase clock error in the probing subroutine, we actually do not know how much information is leaked by the protocol, though we suspect it to be limited. We designate this as outside of the scope of this work and only make claims about privacy when the protocol succeeds. Note that it is impossible to achieve information-theoretic privacy with probability 1 in asynchronous distributed systems because there is always the possibility of premature termination due to indefinite exclusion of agents from the protocol.

\subsubsection{Conclusion}

In this work, we offer various new security definitions in population protocols, such as multiple definitions of privacy which accommodate a range of threat models and scheduling assumptions, and a formal definition of secure peer-to-peer communication. We also develop algorithms solving secure pairwise communication in the model and information-theoretically private computation of the \Remainder\ predicate. In order to show that we can achieve information-theoretic privacy (with high probability) for all semilinear predicates, as in \cite{DFGR2007}, similar algorithms for computing \Threshold\ and \Or\ are also needed. We leave these problems as open for future work.

\bibliographystyle{style/splncs04}
\renewcommand{\appendixbibliographystyle}{style/splncs04}
\bibliography{ref.bib}

\end{document}